# The Static and Dynamic Lattice Changes Induced by Hydrogen Adsorption on NiAl(110)

Aubrey T. Hanbicki[*],
*Naval Research Laboratory, Washington, DC 20375*

P. J. Rous
*Univ. of Maryland Baltimore County, Baltimore, MD 21250*

E. W. Plummer
*The Univ. of Tennessee, Knoxville, TN 37996 and
Oak Ridge National Labs, Oak Ridge, TN 37831*

Abstract

Static and dynamic changes induced by adsorption of atomic hydrogen on the NiAl(110) lattice at 130 K have been examined as a function of adsorbate coverage. Adsorbed hydrogen exists in three distinct phases. At low coverages the hydrogen is itinerant because of quantum tunneling between sites and exhibits no observable vibrational modes. Between 0.4 ML and 0.6 ML, substrate mediated interactions produce an ordered superstructure with c(2×2) symmetry, and at higher coverages, hydrogen exists as a disordered lattice gas. This picture of how hydrogen interacts with NiAl(110) is developed from our data and compared to current theoretical predictions.

[*]Corresponding author:    Aubrey T. Hanbicki
Fax: +202 767 1697
Email: Hanbicki@anvil.nrl.navy.mil

# I. INTRODUCTION

The interaction of hydrogen with the NiAl(110) surface is a prototypical adsorbate-bimetallic ordered alloy system: the surface maintains the bulk stoichiometry and has two components, each with very different chemical properties. Ni and Al are so different in fact, that hydrogen will spontaneously dissociate and adsorb to Ni while there is an activation barrier of ~1 eV over Al [1]. NiAl(110) therefore, provides an ideal test of the concepts associated with whether the global properties of the alloy or the local properties of the individual constituents dictate the overall chemical behavior. Other attributes which make NiAl a particularly attractive prototype system include the facts that it has no major surface reconstructions, it is relatively unreactive, and the clean surface is easy to prepare, maintain and has been studied extensively.

The fact that bimetallic alloys play a significant role in real world catalysis [2] has contributed to growing amounts of experimental information about alloys and multicomponent surfaces. The practical implications, along with recent advances in the theoretical treatment of alloys, have also served to drive theoretical studies of alloy surfaces. In fact, while not without some inconsistency, there has been good success coupling first principles theory with experiment for H/NiAl(110) [3, 4]. This system has even led to a new picture for the nature of activation barriers for dissociative adsorption [5].

The clean NiAl(110) surface has been extensively studied and its physical properties are well known [6-15] making it ideal for the study of alloy surface chemistry. NiAl crystallizes in a CsCl structure, two interpenetrating simple cubic lattices of each atomic species. The (110) termination also has identical interpenetrating unit cells of each atom type and therefore presents equal numbers of both Ni and Al to incoming gases (Fig. 1). This surface can also be thought of as having alternating Ni and Al rows. Numerous structural analyses have been performed and it is well known that the rows of Ni are contracted toward the bulk by 4.0% of the bulk layer spacing, while the rows of Al are expanded out toward the vacuum by 5.5% giving a large static surface ripple (~0.2 Å) [6-9].

The surface vibrational structure of this alloy is also well studied and is correlated with the static ripple [10-12]. Two of the main surface vibrational features are an acoustic surface resonance at 19 meV and an optical surface phonon at 27 meV. For the resonance, the Ni and Al atoms are vibrating in-phase whereas the phonon has the Ni and Al atoms vibrating out-of-phase. The energy and presumably the intensity of the phonon is coupled to the magnitude of the static ripple [11], therefore changes in the surface vibrations gives indications as to the nature of any structural change.

In NiAl, the electronic band structure has been measured and calculated both for the bulk and at the surface [13, 14]. To understand the origin of the barrier toward spontaneous dissociation of hydrogen, it is important that the electronic structure is well known. The accepted view was after Harris and Andersson [16]. In studies based on single component surfaces, they determined that the density of *d*-like states at the Fermi level was the decisive parameter for spontaneous dissociation. A more quantitative calculation by Hammer and Scheffler which considered the NiAl(110) alloy surface, however, presented a model where the dissociative adsorption of molecules at metal surfaces is closely connected to the *depth* of the *d*-band below the Fermi level [5]. Another result of this calculation was that local electronic effects are important. For an alloy, reactions occur on a surface whose electronic properties are globally modified by the alloy. Therefore, experimentally it may appear that global alloy properties control the chemical interaction with hydrogen, however this calculation showed that, while the surface does not behave as if it were a collection of Ni and Al atoms, neither do adsorbates interact with only the global properties of the bulk alloy.

In previous work [3], experimental evidence made it apparent that the NiAl(110) surface behaves very differently from either Ni or Al. We reported that molecular hydrogen has an activation barrier for spontaneous dissociation, with the height of this barrier being 0.72 eV as measured by Beutl et al. [17]. Atomic hydrogen can be chemisorbed to the (110) surface; it desorbs with second order kinetics and a desorption energy of 0.54 eV [3]. The *absolute* coverage of hydrogen on the surface was determined by nuclear reaction analysis with saturation found to be 1 monolayer (ML) of hydrogen [9]. The convention established for this surface was that 1 ML is the number of hydrogen atoms per surface unit cell, not per surface atom as is used with monatomic systems. This density of hydrogen at saturation ($0.85 \times 10^{15}$ H-atoms/cm$^2$) is significantly lower than for the close packed (111) faces of Ni and Al (1.86 [18] and 1.83 [19] $\times 10^{15}$ H-atoms/cm$^2$ respectively).

The symmetry of the surface (Fig. 2) is (1×1) for almost all coverages, except from 0.4 ML to 0.6 ML, where a very weak c(2×2) symmetry is observed [3, 20]. This new symmetry was associated with a hydrogen superstructure because the integrated intensity of the fractional order beams are roughly 1% of the integer order beams. Structural analyses at selected coverages showed that the magnitude of the static surface ripple was reduced but not removed, with increasing hydrogen coverage. At 0.5 ML the ripple is ~0.15 Å as determined by a preliminary x-ray scattering study, and at saturation (1 ML) low energy electron diffraction (LEED) reveals the ripple to be 0.11 Å. Qualitatively this trend is consistent with theory [4, 21].

In this paper, by complementing published results, we present one of the most thorough studies of hydrogen on a metal surface, and undoubtedly the most comprehensive study of hydrogen interacting with a single crystal alloy surface. By monitoring changes in the structural and dynamic properties of NiAl(110) as a function of hydrogen coverage we find some very interesting features which leads to a unique picture. Our model is that hydrogen behaves like an effective potential homogeneously across the surface. As a consequence, increased hydrogen coverage results in monotonic changes in all of the measured surface properties. Spectroscopically, hydrogen is almost invisible on its own, distributed uniformly across the surface and above ~1/2 ML bound over the Ni-Ni bridge sites [4, 20]. We can identify three distinct phases (Fig. 2). At low coverages the hydrogen is delocalized along the Ni rows, itinerant because of quantum tunneling. As coverages become greater than 0.4 ML, substrate mediated interactions conspire to produce an ordered c(2×2) array, which quickly gives way with increasing coverage to a disordered lattice gas. The reason hydrogen affects the NiAl substrate so uniformly probably derives from the fact that the heat of formation is -58.8 kJ/mol [22]; i.e. the NiAl lattice is more rigid than either Ni or Al. Finally, as predicted by theory, hydrogen's predilection for the Ni atoms is experimental evidence that local electronic structure of the substrate is important, i.e. while the alloy properties seem to govern the overall interaction of hydrogen with NiAl, hydrogen does have some sensitivity to the surface Ni and Al atoms.

The organization of this paper is as follows. Sec. II describes the experimental details, which include sample preparation and a description of each of the different techniques used in this study. In Sec. III, the results of structural, vibrational and electronic investigations are presented and discussed in terms of our consistent model. The inconsistencies that still exist with theory [4] are also discussed. Conclusions are briefly summarized in Sec. IV

## II. EXPERIMENTAL DETAILS

All the experiments described below were performed in stainless steel vacuum chambers operating at base pressures of at least $2.0 \times 10^{-10}$ T. The sample was routinely cleaned by sputtering with 1 keV neon ions for 20 minutes with subsequent anneals to 1125 K for 10 minutes to restore surface composition and order. The average surface composition for this crystal is similar to the bulk composition which is near stoichiometry [23]. Once clean, the crystal remained free of contamination for 3-4 hours. Sample cleanliness was verified with a variety of spectroscopic techniques. The particular tool used depended on what was available in the experimental chamber being used, with typical ways of monitoring contamination being HREELS, Auger electron spectroscopy, and the appearance of the O $2p$ peak at 6-7 eV below the Fermi level. Surface order was determined by the presence of a sharp (1×1) LEED pattern.

Both molecular and atomic hydrogen were dosed at sample temperatures lower than 130 K. Atomic hydrogen was prepared by flowing $H_2$ through a tube over tungsten heated to ~1800°C. The sample was placed 3–5 cm in front of this doser tube. Because of the uncalibrated efficiency for producing atomic hydrogen, the exposure of atomic hydrogen is reported as arbitrary units of exposure (E), whose value is simply the Langmuir exposure based on the background pressure. Deuterium was used where specific isotopic effects were expected and was treated identically to hydrogen.

The LEED *I-V* data presented here were collected at the same time as a previous study, and the details of the experimental setup are described there [9]. Data were obtained with a video-LEED system. The sample was aligned perpendicular to the electron beam by comparing *I-V* curves of equivalent spots and requiring they have identical profiles. In addition to the same 11 inequivalent integer beam sets as were collected in the previous studies, 5 inequivalent fractional order beams sets [(1/2,1/2), (1/2,3/2), (3/2,1/2), (1/2,5/2), (3/2,5/2)] were profiled. The total energy range was 4620 eV. Data were taken by averaging over all equivalent beams normalized by beam current. The (1,0) beams were retaken at the end of each data run to ensure the sample condition remained constant.

From the measured *I-V* spectra, the structure of the 0.5 ML hydrogen covered surface of NiAl(110) was determined using the automated tensor LEED (ATLEED) method [24]. In the structural search, atomic relaxations of the substrate atoms in the first two bilayers, both parallel and perpendicular to the surface, were permitted. The hydrogen binding site and height were determined and the hydrogen atom was allowed to move away from high symmetry sites, with the appropriate domain averaging of the calculated IV spectra. Thirteen phase shifts were employed. For Ni and Al, the phase shifts were determined from potentials calculated by Moruzzi, who performed a self-consistent electronic band structure calculation for bulk NiAl, and for H a bare coulomb potential was used. The real part of the inner potential was fit to experiment and damping was represented by an imaginary part of the inner potential of -5.0 eV. Debye temperatures were 390 and 720 K for the Ni and Al, respectively. The automated search algorithm was directed by the Pendry *R*-factor ($R_p$) [25].

The work function ($\phi$) and changes in the work function ($\Delta\phi$) as a function of dose were monitored with a photon incident angle of 45°, normal emission, and a negative bias on the sample. A stationary Fermi level was monitored for the first 20 measurements and was completely reproducible; the position of the low kinetic energy cutoff voltage was used to determine the change in the work function. Additional work function measurements, using a retarding field method, were performed in conjunction with coverage measurements [9] and LEED observations. The details of the experimental setup have been described elsewhere [9]. 30 eV electrons from the LEED electron gun illuminated the crystal while sample current was recorded as a function of sample bias. The change in cutoff voltage, i.e.

the voltage where the current dropped to zero, was monitored as a function of absolute coverage and LEED symmetry.

Vibrational information was collected with commercial LK-2000 and HIB-1000 HREELS spectrometers. The LK-2000 had typical resolutions of 5-6 meV; it was operated in the specular geometry with the incident and exit angles at 60°; off-specular measurements were performed by rotating the analyzer. Spectra from the HIB-1000 had resolutions of 2.5-3.5 meV; the incident and exit angles were 45°; off-specular measurements were performed by rotating the monochromator. Full azimuthal rotation of the sample was possible, and specular and off-specular data were taken with energies ranging from 1.5 eV to 136 eV.

ARUPS experiments were carried out at beamline U-12 of the National Synchrotron Light Source at Brookhaven National Laboratory. The beamline is equipped with a toroidal grating monochromator and an angle-resolved hemispherical electron-energy analyzer [26]. Before reaching the sample, the light traverses a tungsten mesh providing a normalization current of the incident flux from the monochromator. The sample was mounted with its high symmetry directions within 2° of the linearly polarized beam, the light being polarized parallel to the $(1\bar{1}0)$ direction.

## III. RESULTS AND DISCUSSION

*A. Surface Structure:*

A full, dynamical LEED *I-V* structural analysis has been reported on this surface for both the clean and saturated (1 ML) surface at 130 K [9]. It was found that the large ripple in the NiAl(110) surface was lessened, but not completely removed, upon hydrogen adsorption. A preliminary x-ray analysis [9] performed at 0.5 ML provided a structural determination at a coverage intermediate to the LEED analyses and indicated the reduction of the ripple occurs smoothly. Because of the limited data set collected in this preliminary x-ray analysis, however, it could not provide the level of structural detail supplied by the LEED analysis reported in this paper. Further, the determination of hydrogen position is impossible with x-ray diffraction. The fractional order beams seen with LEED and associated with a hydrogen superstructure are not visible with x-rays due to the relatively small cross section for x-rays with the surface hydrogen. Therefore, we have performed a complete structural determination of the 0.5 ML hydrogen covered surface using ATLEED, and by inclusion of the fractional order beams, determined the hydrogen bonding site in addition to the substrate structure.

Structure determination began by performing a limited structural search that fit the measured integer-order *I-V* spectra to calculations which neglected the hydrogen overlayer and included only electron scattering from the NiAl(110) substrate. The initial reference structure was the bulk termination of NiAl(110), but as is the standard procedure for ATLEED, new reference structure calculations were performed after the termination of each ATLEED search. Only displacements of the Ni and Al atoms normal to the surface were permitted, preserving the (1×1) symmetry of the unreconstructed substrate. The results of this restricted analysis are presented in table I. Figure 3 shows the measured beams (solid curve) and *I-V* profiles calculated for this structure (dashed curve) for a representative selection of integer order beams. The best fit structure has $\Delta d_{12}(Ni)=-4.0\pm0.7\%$, $\Delta d_{12}(Al)=+3.4\pm1.0\%$, $\Delta d_{23}(Ni)=0.0\pm1.1\%$, and $\Delta d_{23}(Al)=-0.8\pm1.5\%$ with a Pendry *R*-factor $R_p=0.17$. Error bars were obtained using the procedure suggested by Pendry [27] and are slightly smaller for the Ni atoms which are, on average, stronger scatterers than Al. The most significant structural feature is the reduction of the rippling of the first layer from the clean value of 9.4% [9] to 7.2% or from 0.20 Å to 0.15 Å. A slight rippling of the second layer of magnitude 0.8% or 0.02±0.02 Å is also observed. It can be noted here that in the second layer the direction of the rippling is reversed relative to the first layer, a trend also seen for the clean [7, 9] and hydrogen saturated surface [9].

Using this structure as the initial reference surface, a more comprehensive search was launched using an ATLEED calculation that now included the hydrogen scattering. A fit was performed to both the integer and fractional-order *I-V* spectra. Variations of both the lateral and the normal coordinates of the hydrogen atom and the Ni and Al atoms in the first two layers, consistent with the observed c(2×2) symmetry, were now permitted; a total of nine structural parameters were varied. The results of this analysis are presented in table I. Integer order *I-V* profiles calculated from this best-fit structure also appear in figure 3 (dotted line) for comparison to the fit where no hydrogen was considered. Normal coordinates of the substrate atoms are $\Delta d_{12}(Ni)=-3.6\pm0.5\%$, $\Delta d_{12}(Al)=+3.6\pm1.0\%$, $\Delta d_{23}(Ni)=0.0\pm1.1\%$, $\Delta d_{23}(Al)=-0.8\pm1.5\%$. The Pendry *R*-factor, calculated only for the integer order beams, is $R_p=0.18$, almost unaffected by the inclusion of hydrogen. As is clear from figure 3, within the error bars, these parameters agree well with the results of the earlier search, where scattering from hydrogen was neglected. Again, the magnitude of the first layer ripple is 7.2% or 0.15±0.02 Å and the second layer ripple is 0.8% or 0.02±0.03 Å. This is in excellent agreement with the preliminary x-ray result (0.15 Å) [9], as well as first principles theoretical predictions of 0.147 Å for the first layer ripple and 0.06 Å for the second layer where 0.5 ML of hydrogen in a c(2×2) symmetry was considered [21].

The experimental and theoretical beams for the fractional order spots are shown in figure 4. Note that the intensity scale is the same as used for the integer order beams. Because hydrogen is a very weak scatterer with a scattering cross section roughly two orders of magnitude smaller than transition metals like Ni, it is tempting to attribute these very weak fractional beam intensities to hydrogen superstructures. Up to this time, however, for all systems studied with LEED, the intensity in the fractional order beams could always be explained with slight changes in the positions of the substrate atoms rather than with hydrogen superstructures [28, 29]. This includes the Ni(111)-(2×2)-2H system where originally a hydrogen superstructure was identified [18], but a reanalysis has found that in fact slight substrate modification was necessary to reproduce the fractional order beams. Models which considered only a hydrogen superstructure produced fractional order beams which were ~1% of the integer order beams, but the experimental value was ~3% ; the difference could only be made up with substrate reconstruction [30].

The intensity in the experimentally collected fractional order beams is ~1% of that for the integer order beams and can *not* be reproduced with any substrate reconstruction that we tried. In figure 4, the solid line is the experimental data, the dashed line is generated from a fit where only substrate modification was considered, and the dotted line is the best fit, produced mainly from a hydrogen superstructure. It is obvious from this figure that the agreement between theory and experiment in the fractional order beams is significantly better if hydrogen scattering is considered. Quantitatively, the *R*-factor between experiment and beams produced by the substrate modification, calculated only for the fractional order beams, is $R_p$=0.4. This high *R*-factor reflects the inability of lateral reconstructions of the substrate alone to reproduce the observed fractional spectra. Consequently, they must originate from scattering in the hydrogen overlayer. For the hydrogen superstructure, again considering only the fractional order beams, $R_p$=0.2, a value comparable to that of the integer beams. This result suggests the structural origin of the scattering into the fractional beams has been well reproduced by the best fit structure. For our best fit structure, the lateral displacements of all of the Ni and Al atoms away from their location in the bulk termination was found to be smaller than 0.05 Å. Since the error bars in these lateral positions were all greater than 0.06 Å, *we find no evidence for lateral reconstruction of the substrate* induced by hydrogen adsorption.

The fractional beams, therefore, provide information on where the hydrogen is bound. Table II gives $R_p$ for hydrogen in several different bonding sites. From this data it is clear that hydrogen is located in the Ni-Ni bridge. The exact location of the hydrogen is fairly poorly determined, however, with a best fit yielding the hydrogen height to be 0.6±0.8 Å above the Ni plane. The error on the height of the hydrogen is so large because the error bars are derived from the *R*-factors achieved using all of the beams (integer as well as fractional). Because the integer order beams are almost completely insensitive to the hydrogen, the error bars are correspondingly large. The error in the hydrogen height is reduced if only the fractional order beams are considered. In this case, the error is ±0.21 Å. This position corresponds to a Ni-H bond length of 1.56 Å and a hydrogen radius (the difference between the observed bond length and the metallic radius of the substrate atom) of ~0.35±0.08 Å. This is consistent with other structural studies of hydrogen adsorption at metallic surfaces using LEED, for instance one study found an hydrogen radius of 0.49±0.08 Å on the close packed surface of Ni and 0.58±0.2 Å on the close packed surface of Fe [30], while another study finds the hydrogen radius to be 0.41 Å when adsorbed to Be(0001) [29]. A first principles calculation for the hydrogen bonding site is also the Ni-Ni bridge for both 0.5 ML [21] and 1 ML [3], however the hydrogen bond length is significantly longer than is seen experimentally, with the hydrogen being roughly 0.95 Å above the surface. This large hydrogen radius of 0.7 Å could be a consequence of allowing only the top layer to relax. Theoretical calculations by Konopka, et al. [4] predict that, for 1 ML of hydrogen, the bonding site is tilted slightly out of the Ni-Ni bridge. In this theory, the bond length is a reasonable 0.52 Å.

Our model, presented in the introduction section, has its roots in these ATLEED analysis results reported here. Even though hydrogen is identified as being in Ni-Ni bridge sites at this coverage, no new symmetry is introduced to the substrate. Rather, the ripple is reduced evenly over the surface, being influenced by the effective potential which is the hydrogen.

### B. Work Function vs. Coverage:

The origin of the work function for a solid has both a surface contribution, the formation of a surface dipole layer, and a bulk contribution, the chemical potential. Changes in the work function can be caused by either structural distortions which lead to surface charge rearrangement, or simply electronic redistribution caused by adsorbed species. In some cases it is possible to isolate which of these effects is responsible for shifts of the work function. Changes in the work function induced by hydrogen adsorption on a metal surface can be usually understood in a simple way. The electrostatic dipole associated with the surface barrier is modified as a result of the hydrogen-metal bond; because hydrogen has a high electronegativity, there is a net charge transfer to the hydrogen. Indeed, consistent with this picture the trend for most transition and noble metals is to increase the work function as hydrogen is adsorbed [31].

A few notable exceptions from this trend are H/Fe(110) [32], H/Pt(111) [33], and H/W(110) [34] where the work function actually decreases. A good illustration of the deviations possible from the simple model of charge transfer is a comparison of the Ni(111)-(2×2)-2H system where Δϕ is positive, and the Fe(110)-(2×2)-2H system where the work function decreases. The adsorption geometry on both of these surfaces is identical, however, the substrate modification is fundamentally different: the Ni atoms are buckled toward the hydrogen, while the Fe atoms are buckled away. Therefore, a redistribution of substrate charge resulting from a surface rearrangement may dominate Δϕ.

The clean work function for NiAl(110) is determined to be 4.81±0.04 eV from photoemission, in good agreement with previous work [14]. This is roughly the average of the work functions for clean Ni(111) and clean Al(111) which are 5.35 eV [35] and 4.25 eV [36] respectively. Hydrogen adsorption to NiAl(110) *decreases* the work function monotonically as a function of coverage (Fig. 5). For coverages between 0.4 ML and 0.6 ML, i.e. for the c(2×2) structure, Δϕ is between -0.12±0.02 eV and -0.22±0.02 eV. At saturation, the work function is reduced by 0.6±0.05 eV.

At the highest coverages, where contamination is a problem, the work function shift can be much larger than -0.6 eV. This is similar to an effect noted during absolute coverage measurements: below 130 K, $H_2O$ will adsorb on NiAl(110) [37], so coverages greater than 1 ML are achievable for high exposures due to the adsorption of water [9]. The work function shift induced by water adsorption on metal surfaces is usually on the order of -1.0 eV [38]. Thus as the large doses required to achieve higher coverages lead to contamination and coverages greater than 1 ML, so to can spuriously high changes in the work function be expected at these exposures. Indeed, an abrupt change in the slope of Δϕ at 0.9 ML is clearly visible in figure 5. Likely this is a result of small amounts of adsorbed water.

As discussed, a work function decrease means that hydrogen is not adsorbing to this surface in a simple way. Any model must include either the population of subsurface sites or a hydrogen induced restructuring of the surface, e.g. an interplanar shift or surface buckling and hence a surface charge redistribution. The latter of these possibilities is immediately suggested by the structural studies, i.e. the change in the surface dipole must be dominated by the change in the surface charge corrugation associated with the decrease in the surface ripple. A combination of subsurface hydrogen and reconstruction effects is not ruled out either, though subsurface hydrogen is unlikely for several reasons. First, in general there is always a surface species of hydrogen on metal surfaces. Even in such cases as H/N(111) [39] where subsurface hydrogen has been identified, it is preceded by a surface species. Also, a smooth change in work function suggests that hydrogen is populating only one plane, e.g. the surface or a single subsurface layer. If a switch from surface to subsurface hydrogen were to occur at some coverage, a kink in the slope of $\Delta\phi$ should be expected because of a change in the sign of the surface dipole. Such a kink does exist, though not until 0.9 ML when contamination is an issue. Coupled with the facts that a single peak is seen with thermal desorption spectroscopy [3], suggesting a single bonding site, the LEED analysis described in the preceding section which finds the hydrogen bound in the Ni-Ni bridge at 0.5 ML, and absolute coverage measurements which show saturation to be 1 ML [9], consistent with one full layer of hydrogen, subsurface hydrogen does not seem to be an option.

With the possibility of subsurface hydrogen excluded, the negative change in work function must therefore be related to the change in surface structure. More specifically, the change in substrate ripple dominates the change in work function. Therefore, the smooth decrease in the work function means the two structural data points can be extended to include all of coverage space, i.e. *the surface ripple decreases monotonically with coverage*.

First principles calculations also predict a decrease in work function with increasing hydrogen coverage, but the magnitude of change is somewhat larger than the observed values: -0.46 eV at 0.5 ML and -1.06 eV at saturation [21]. Because the work function is dominated by the ripple, such a result might be expected. Again, by not considering the structural changes in the deeper layers, the decrease in surface ripple, and by extension the work function change, could be overestimated by theory. It may also be a sign of a more general problem with the LDA, however. Similar calculations by Feibelman and Hamann for hydrogen on Pt(111) also predict a large H-Pt bond length, and overestimate the decrease in work function as compared to experiment [40].

The work function data can now be used to further develop the model initiated with the ATLEED results. As hydrogen adsorbs on the surface it behaves as an effective potential. The surface properties, e.g. the rippled structure, are affected monotonically as a function of coverage for all of the coverages considered here.

### C. Vibrational Structure:

While the data presented above provides information on the static properties of the interaction of hydrogen with NiAl(110), a full picture can only be understood with the inclusion of dynamic properties (vibrations) and electronic structure. On clean NiAl(110) there are two main surface vibrational features, a surface phonon at 27 meV and a resonance at 18-19 meV [11, 12]. For an alloy system like NiAl, the large difference in mass between the constituent atoms results in an appreciable energy gap between the bulk vibrational optical and acoustic branches. The 27 meV mode is a surface phonon which lies in the middle of this bulk vibrational gap, split off from the bottom of the optical branch. Because it exists where there are no bulk states, the phonon is localized in the surface region, decaying exponentially into the bulk. At the center of the surface Brillouin zone, the motion of this optical vibration has the surface Ni and Al atoms vibrating out-of-phase with amplitudes (u) related by $u_{Ni} = -0.4\ u_{Al}$ [10]. Calculations for a clean, bulk truncated surface have the phonon at 29 meV [11]. Because the surface is rippled however, and more specifically because the Al atoms are relaxed toward the vacuum, the surface Al force constant becomes smaller than its bulk value and the energy of the phonon is decreased to 27 meV [4, 10]. In contrast, a surface resonance mode lies within the bulk acoustic band and the surface Ni and Al atoms vibrate in-phase (at 20 meV) [4, 11] with roughly the same amplitude ($u_{Ni} = 1.1\ u_{Al}$) [10].

Figure 6 displays a series of specular ($q_{\parallel}=0$) HREEL spectra with increasing hydrogen coverage. The shaded regions on this figure are the bulk vibrational bands projected on to the (110) surface. Increasing hydrogen coverage affects the energy and cross section of the surface phonon, the surface resonance and a new vibration identified as a hydrogen mode. Consider first the behavior of the surface phonon, indicated in figure 7 by filled circles. The phonon energy (Fig. 7a) increases monotonically with hydrogen coverage to 33.25±0.15 meV, just below the bottom of the bulk optical band (35.6 meV) at saturation. At 0.5 ML coverage when the c(2×2) LEED pattern is visible, the phonon energy is 29±1 meV. Theoretical calculations by Kang and Mele for the clean surface predict the phonon to be at 27 meV [10], and calculations by Hammer which include 0.5 ML hydrogen predict a phonon energy of 31 meV [21]. The intrinsic linewidth of the surface phonon can be determined by fitting it with a Lorentzian line shape convolved with the elastic peak to remove the instrumental response from the measured signal [41]. For all coverages, the phonon has a constant width of 1.95±0.4 meV (Fig. 7b). This is comparable to widths found in other EELS studies, for instance Cu (2.1 meV) [42] and Be (1.4 meV) [43]. Finally, the cross section of the phonon, represented by the peak area, increases nonlinearly with hydrogen adsorption (Fig 7c).

The monotonic shift of the phonon energy suggests a smooth change in surface structure consistent with conclusions drawn from the LEED and work function data. An indication of the nature of the structural change can be understood in the context of the force constants already discussed. Just as the relaxation of the surface Al atom in the clean, rippled surface leads to a lower phonon energy and a corresponding decrease in the surface Al force constant, in a simple model an increase in energy of the phonon means the surface region Al force constant is increasing. Along with this increased force constant is likely a relaxation of the Al atoms back toward their bulk terminated positions. This is consistent with the two structural data points already discussed and corroborates the observation that the surface ripple is being reduced. Just as with the work function data, because the phonon change is smooth throughout the coverage regime, it is possible to extend the observation provided by the structural determination and conclude that *with increasing coverage the surface ripple is reduced monotonically and homogeneously with increasing hydrogen coverage*. The phonon energy shift is therefore completely consistent with the conclusions drawn earlier from the change in static properties.

While the phonon energy shift corroborates the previous observations, additional information can be drawn from a constant inherent phonon linewidth. There are two main components to the linewidth, a coherent part, the intrinsic decay of the phonon, and an incoherent part which includes surface defects and roughness. Assuming the intrinsic component will not change as a function of coverage, a constant width has two implications. First, no inhomogeneous broadening occurs; that is to say, hydrogen does not form islands but rather is distributed homogeneously across the surface. Rather than affecting local sections of the surface, hydrogen causes the reduction of the surface ripple to occur everywhere on the surface at the same time. Second, the surface roughness remains constant, i.e. hydrogen does not nucleate defects nor does it increase step mobility which may cause decreased terrace sizes. Instead, the roughness in the surface after a cleaning cycle remains unaffected by hydrogen adsorption. Perhaps more convincing evidence of this is provided by the integrated intensity of the elastic peak just before and after hydrogen adsorption. The top curve in figure 7c (represented by the + symbol and referring to the axis on the right) is the ratio of the elastic peak with and without hydrogen. If significant roughness were introduced on the surface, the elastic intensity after the adsorption of hydrogen would be noticeably reduced. For hydrogen on NiAl(110), however, the elastic intensity actually *increases* after exposure to atomic hydrogen. Changes in the spectrometer tuning could also take place when exposed to hydrogen, but the overall trend is clear, hydrogen has no local effects on the substrate.

The phonon width in the 0.5 ML coverage regime with the c(2×2) symmetry requires some additional attention. With this new surface symmetry, the $\bar{S}$ point becomes a new zone center (see Fig. 1) and any modes that exist there are folded back to $\bar{\Gamma}$. For the clean surface, however, we did not observe any new features which could be folded back from the new zone boundary. Therefore, because the c(2×2) region doesn't contain any new vibrational structure due to the clean surface, and because the phonon has a constant width for all coverages, there is no vibrational signature of the c(2×2) coverage regime. The hydrogen mode at 49.5 meV begins concomitantly with the c(2×2), but it persists even after the surface symmetry returns to (1×1). This confirms the conclusions drawn from the ATLEED analysis: the substrate is not more ordered in the 0.4-0.6 ML coverage regime even though a new symmetry is present. The hydrogen forms an ordered superstructure and has no major effect on the overall order of the NiAl.

The phonon's increase in cross section (Fig. 7c) is data which is counterintuitive. Again, the phonon consists of the Ni and Al atoms vibrating out-of-phase. Because the phonon is a dipole active peak, the cross section might be expected to decrease as the surface ripple decreases because the Ni and Al atoms get closer together, reducing the strength of the dipole. Chen et al. have performed a detailed electron energy loss cross section analysis of the clean NiAl(110) surface in an attempt to obtain a unique lattice-dynamical model. Their calculations show that the cross sections are sensitive functions of the surface structure [44]. Even when the correct surface dynamical model is used, i.e. if the right force constants are employed, if the wrong surface geometry is considered it is no longer possible to get good agreement between experiment and theory. Therefore, to fully understand the cross section of the phonon as a function of coverage, calculations must be performed which consider the real surface structure . Like the phonon, the cross section of the surface resonance also increases with coverage (triangles in Fig 7c). Considering the arguments made for the phonon, this too can be expected because of the cross sections geometry dependence [44].

Thus far, the change in energy and width of the phonon has served to further develop and confirm the conclusions made from the structural and work function studies. As hydrogen adsorbs, the entire surface is being affected in the same way, i.e. the hydrogen is distributed evenly over the entire surface acting as an effective potential reducing the surface ripple, and subsequently the work function, in a smooth way. Examination of the induced adsorbate vibration will help to complete the picture. This hydrogen mode turns out to be a source of glaring disagreement between experiment and existing theory.

For hydrogen localized in the Ni-Ni bridge, theoretical predictions claim there should be a well defined hydrogen vibration with an energy of about 135 meV, calculated both within [21] and independent of the harmonic approximation[4]. This is certainly not the case (Fig. 6b). Attributes of the hydrogen mode are plotted in figure 7, and represented by the open squares. Hydrogen induces an adsorbate vibration on the NiAl(110) surface at $49.5\pm0.2$ meV only for coverages greater than ~1/2 ML. This peak decreases slightly in energy (Fig. 7a) with increasing coverage and is attributable to a hydrogen mode because it exhibits an isotopic shift. The inherent width (Fig. 7b) remains constant at $2.4\pm0.4$ meV, the intensity (Fig. 7c) however, increases more rapidly than linear with coverage (Fig. 7c). Figure 8 shows HREEL spectra proceeding away from the specular direction for a coverage of 0.61 ML. The rapid decrease in intensity of the hydrogen mode means it is dipole active. It is important to note that upon hydrogen adsorption, there are *no* vibrations other than those associated with the substrate and the 49.5 meV peak either specular or off-specular. At high exposures many additional adsorbate vibrations can be seen, however, published studies of the adsorption of oxygen, carbon monoxide, water, and methanol [17, 23, 37, 45-47] make identification of contamination straightforward.

The behavior of the adsorbed hydrogen is very unusual and has two distinguishing features. First, for hydrogen coverage less than 1/2 ML, no adsorbate induced vibrational modes are observed either specular or off-specular. Second, when coverages become greater than 1/2 ML, a new vibrational feature does evolve, but at 49.5 meV. Hydrogen on metal surfaces typically induce adsorbate vibrations with energies ranging from 75 meV to 150 meV [31], the higher the coordination the lower the energy. The energy observed here is anomalously low.

One possible explanation for the lack of a hydrogen vibration could be that, at first, hydrogen populates subsurface sites, leading to no adsorbate vibrational intensity. After 0.5 ML the entrance to subsurface sites could be blocked and the hydrogen starts to then adsorb to the surface, giving a measurable HREELS signal. This is not the case, as has been argued in light of the work function data. In fact, subsurface hydrogen can be seen with vibrational spectroscopy and has been identified for hydrogen embedded in Ni(111) [39], showing that intensity from the subsurface species is possible. A minor caveat is that the density of subsurface hydrogen leading to the HREELS signal for the H/Ni(111) system was roughly 5 times greater than the density of hydrogen in the low coverage regime of this study.

A more likely explanation is that, at the lowest coverages, the hydrogen is delocalized, itinerant along the Ni rows (Fig. 9a). The idea that hydrogen can be delocalized on the surface of a metal was originally proposed by Christmann et al. [18], and has since been suggested for several systems [48-52], including a recent helium atom scattering work performed by Farias et al. [20]. Delocalized quantum motion of hydrogen is very similar to a free electron gas system. Hydrogen may be strongly localized perpendicular to the surface, however, for systems where the surface corrugation is small, it may be completely delocalized parallel to the surface. A rigorous treatment of such a system has been presented for hydrogen on the three low index faces of Ni by Puska et al. [53, 54]. In that study, the starting point was to calculate a potential energy surface for hydrogen outside a low index Ni surface. In doing so, they found a large anharmonicity which leads to a strong coupling between motion parallel and perpendicular to the surface. Because the potential is no longer separable there does not have to be a mode associated with just the vertical localization. For a proper description of the ground state and the excitation spectra it is necessary to solve the full 3-D Schrödinger equation. Such a treatment leads to protonic vibrational *bands*, analogous to the electronic bands formed by the conduction electrons of a simple metal. Hydrogen is most delocalized in the close-packed directions, where the corrugation is the smallest, and leads to vibrational bands with widths of ~20-40 meV. The lack of hydrogen signature in HREELS is therefore a convolution of effects. The density of hydrogen is very low (remember below 1/2 ML for this surface is less than 1 H per 4 surface atoms), which leads to a very weak signal. This is coupled to the fact that the intensity in this weak vibration is spread out over a wide energy range, including some of the bulk vibrational continuum, effectively making the vibration unobservable. This concept of invisible, delocalized hydrogen on the surface is consistent with our model of hydrogen as an effective potential interacting homogeneously over the entire surface.

What then occurs to make a hydrogen vibration appear at higher coverages? When the hydrogen reaches 0.5 ML, a substrate mediated H-H interaction localizes the hydrogen into sites providing the largest distance between each atom. This new geometry has a c(2×2) symmetry (Fig. 9b). This conclusion is borne out by the ATLEED study where we found that the fractional order spots in the c(2×2) coverage regime are in fact caused by a hydrogen superstructure rather than a restructuring of the substrate to a new symmetry. The hydrogen vibrational energy is, therefore, now well defined because the hydrogen is localized in Ni-Ni bridge sites.

In the coverage regime greater than 1/2 ML, several questions remain to be reconciled with the experimental observations: why does the hydrogen mode have such a low energy, why is the intensity increase greater than linear, and why does the hydrogen superstructure disorder to a lattice gas above ~0.6 ML (Fig 9c). Turning first to the low energy of the hydrogen vibration, if the potential energy surface (PES) of the adsorbed hydrogen is not strictly harmonic, but rather a 3 dimensional *non*-separable PES, the in-plane and out-of-plane motion will be coupled and significantly lower vibrational energies will be predicted. For the case of H/Ni this has been calculated [53] and the hydrogen vibration was predicted to be at 62 meV rather than the 76 meV predicted with an harmonic potential, a decrease of 18%. A harmonic PES on NiAl(110) leads to a theoretical prediction of 135 meV [4, 21], a value 63% greater than that observed: can a hydrogen mode can occur at energies previously considered too low if the simple harmonic approximation is invalid. Konopka, et al. claim that while the 135 meV should exist, it may not be dipole active, and therefore won't be seen experimentally. They also speculate that a mode at 50 meV can be caused by impact scattering from an in-plane vibrational mode [4].

After 1/2 ML the intensity increase is greater than linear with coverage. An interesting possibility arises if the surface image plane is located directly between the first layer Ni and Al atoms. Assuming the hydrogen is always over the Ni with a fixed Ni-H separation, as hydrogen coverage increases and surface reconstruction is induced, the hydrogen-image plane distance will increase. Thus would the transition dipole component normal to the surface increase. Therefore, the intensity increase of the hydrogen vibration increases faster than linearly because not only is the amount of hydrogen increasing, but it is also modifying the hydrogen to image plane distance. This trend continues until saturation at 1 ML (Fig. 9d).

Finally, turning to the question of hydrogen order, at ~0.5 ML the hydrogen forms an ordered overlayer. As hydrogen coverage increases, the hydrogen does not island in regions where the ordered overlayer remains, rather the LEED pattern returns to a (1×1) indicating the hydrogen forms a lattice gas, i.e. it has disordered on the surface. Now though, instead of the hydrogen being delocalized, the presence of an adsorbate vibration indicate it remains in the localized Ni-Ni bridge sites. To understand this behavior, the nature of the substrate mediated hydrogen-hydrogen interaction must be considered. Because hydrogen adsorbates represent impurities on the surface, they will be screened by the surface electrons. Surface Friedel oscillations will occur which decay rapidly away from the point defects (Fig. 10). Such surface waves have been seen on the NiAl(110) surface [55]. At some coverage, the hydrogen will be close enough that the phases between these oscillations will interfere [56]. An ordered c(2×2) structure can form if at 0.5 ML all of the phases add constructively. Such constructive interference will occur only for a small coverage regime so for coverages greater than ~0.6 ML substrate mediated effects will therefore be diminished; the hydrogen will no longer "see" each other and be disordered. Just such as scenario has been directly observed for Sn defects on a Ge(111) substrate [57]. Using STM, Melechko, *et al.* have seen adsorbates interact through defect density waves.

*D. Electronic Structure:*

The NiAl(110) surface has many surface states which are well-documented [13]. Because of the geometry of the experiment, the surface states with $\Sigma_1$ and $\Sigma_4$ symmetries could be seen, however, the surface state with $\Sigma_3$ symmetry was not accessible. The $\Sigma_4$ surface state was observed to change as a function of hydrogen coverage. This surface state has a maximum in intensity at 17 eV and was monitored with an incident angle of 15° (nearly *s*-polarized light) and normal emission ($\bar{\Gamma}$). Figure 11 clearly shows that as the coverage of H is increased, the surface state disappears. The intensity decreases while the surface state shifts to higher binding energy from its clean surface value of -1.10±0.05 eV, quickly merging with the *d*-band.

With a photon energy of 35 eV and a 45° incident angle (*p*-polarized light), at high doses of hydrogen, a new peak appears in the spectrum at $\bar{Y}$. Figure 12 shows two energy distribution curves (EDCs), a clean spectrum and an H saturated spectrum. The inset on this figure is the difference between them, showing the new feature with a binding energy of -7.26 eV, and a width of 1.71 eV. The dispersion of this peak was not measurable because the NiAl feature at slightly lower energy visible in the clean EDC, effectively masks any shifting.

Observations of changes in the electronic structure of NiAl(110) as a function of adsorbed hydrogen are hardly surprising. Because electronic states confined to the surface are extremely sensitive to adsorbates, the removal of any surface states on NiAl(110) should be expected. The origin of the $\Sigma_4(\Sigma_3)$ surface state is a distortion of the d-like charge lobes extending toward the Ni-Ni bridge in the [110]([001]) direction [13] due to the loss of symmetry created by the surface. Chemisorbed hydrogen, with the accompanying charge transfer and surface modification will reduce or eliminate all clean surface charge rearrangements, and indeed, the $\Sigma_4$ surface state disappears with coverage. Presumably the $\Sigma_3$ surface state also suffers modification.

The appearance of a hydrogen bonding state is also expected as there are many other systems where a new bonding feature due to hydrogen is observable in the 5.5-10 eV range [58]. These hydrogen induced bonding features are far enough below the Fermi level that they are interpreted as the bonding level of hydrogen formed by interaction with the metal *s* electrons. The energy of -7.26 eV below the Fermi level is in decent agreement with the calculation provided by Hammer [5]. While not specifically presenting the hydrogen bonding level, his calculations show a bonding feature between the hydrogen molecular bonding and antibonding levels and the metal substrate *s* electrons to be in a range between 6-8 eV below the Fermi level when the molecule is on the surface.

## IV. CONCLUSIONS

The exothermic heat of formation for NiAl, and the stability of the (110) surface make this a unique hydrogen-on-metal system. From all of the experimental evidence, it is apparent that hydrogen behaves like an effective medium on the NiAl(110) surface. Increased hydrogen coverage homogeneously changes all of the measured surface properties in a monotonic way. This includes a smooth reduction in the surface ripple from 0.19 Å to 0.11 Å, a reduction in the work function by 0.6 eV, an increase in the surface phonon energy from 27 meV to 33 meV without any change in the inherent linewidth, and a removal of the $\Sigma_4$ surface state. Hydrogen is almost invisible on its own, distributed uniformly across the surface and bound in very anharmonic potentials at the Ni-Ni bridge sites. With increasing coverage, the adsorbed hydrogen proceeds through three different phases. At low coverages the hydrogen is delocalized, probably along the Ni rows, invisible to vibrational spectroscopy and itinerant because of quantum tunneling. As coverages become greater than 0.4 ML, substrate mediated interactions localize the hydrogen into an ordered c(2×2) array. This phase is the first observed hydrogen overlayer where the substrate doesn't reconstruct with the same symmetry. Greater than 0.6 ML, the adsorbed hydrogen disorders into a lattice gas. The qualitative agreement with theoretical calculations for all aspects of this interaction is remarkable.

On a fundamental level, this system has provided an ideal test of the concepts associated with whether the global properties of the alloy or the local properties of the individual constituents dictate the overall chemical behavior. Experimentally, the global alloy properties seem responsible for interactions. The activation barrier to the spontaneous dissociation of molecular hydrogen is the most obvious consequence, but in the model presented here, the fact that hydrogen behaves as an effective potential is also a manifestation of this fact. One major prediction by theory is that local electronic structure of the substrate is important. Hydrogen's inclination for the Ni atoms, at first delocalized along the Ni rows and then localized at Ni-Ni bridge sites, is experimental evidence that this is in fact the case. It is important to note, however, that the alloy modifications to the surface Ni atoms make them very different from pure Ni atoms. A further understanding of how hydrogen interacts with the NiAl alloy, and more generally an understanding of what affect minor changes in local electronic structure have toward the interaction of adsorbates with metal surfaces can be derived from studies of hydrogen interacting with the other low index faces of NiAl.

## ACKNOWLEDGMENTS

We would like to thank Bjørk Hammer for providing us with theoretical results and David Zehner who was instrumental in making this study possible. Funding for this work was provided by NSF-DMR 0105232. A portion of this work was conducted at Oak Ridge National Laboratory, managed by UT-Battelle, LLC, for the U.S. Department of Energy under contract No. DE-AC05-00OR22725.

**Figure Captions**

Figure 1---The NiAl(110) surface structure is shown. a) top view, b) side view showing the ripple in the surface; c) reciprocal space (1×1) and c(2×2) unit cells with the appropriate high symmetry points labeled.

Figure 2---Phase diagram of surface symmetries induced by the adsorption of hydrogen as a function of coverage and temperature.

Figure 3---Six inequivalent experimentally observed integer order *I-V* profiles (solid line) compared with those generated from the best fit model where hydrogen was **not** considered in the model (dashed line), and from the best fit model where hydrogen **was** considered (dotted line). The curves have the same intensity scale, but have been offset for clarity.

Figure 4---*I-V* profiles of the fractional order beams. The solid line is the experimental data, the dashed line is generated from a model which considered only substrate modification, and the dotted lines are generated from a hydrogen superstructure. Note that while each of these curves has been offset for clarity, the intensity scale is commensurate with that of the integer order beams.

Figure 5---Change in work function as a function of absolute hydrogen coverage. The gray area indicates the region where the LEED symmetry is c(2×2).

Figure 6---Series of high resolution specular vibrational spectra. For clarity, each of the curves is offset. The gray bands are the projection of the bulk vibrational bands on to the (110) surface.

Figure 7---Parameters of various peaks seen with vibrational spectroscopy in the specular geometry vs. absolute hydrogen coverage. a) Energy of the resonance (triangles), phonon (filled circles) and hydrogen mode (squares) The projection of the bulk vibrational bands is indicated on the right axis. b) Width of the phonon (filled circles) and hydrogen mode (squares). c) Peak area for the resonance (triangles), phonon (filled circles) and hydrogen mode (squares). The peak areas have been offset for clarity. In c), the axis on the right refers to the top curve (crosses) and is the ratio of the elastic peak with and without hydrogen.

Figure 8---Off-specular vibrational spectra with 0.61 ML of hydrogen adsorbed.  The inset shows the direction and distance being probed in reciprocal space.

Figure 9---Ball model of the hydrogen on NiAl system for different coverage regimes.  a) Low coverage where hydrogen is delocalized on the Ni rows; b) 0.5 ML with the hydrogen localized in Ni-Ni bridges forming a c(2x2) superstructure; c) 3/4 coverage and the hydrogen, while still in Ni-Ni bridges are randomly distributed; d) saturation, 1 ML and all of the available adsorption sites are filled.

Figure 10---a) Schematic of a Friedel oscillation resulting from a point defect on a surface; b) Friedel oscillations adding constructively so that minima will occur in a c(2×2) superstructure.

Figure 11---Series of EDCs with increasing hydrogen exposure.  The $\Sigma_4$ surface state is indicated.

Figure 12---EDC from a clean NiAl(110) surface (thin line) and that from a surface saturated with hydrogen (thick line).  The hydrogen saturated spectrum is a group of 5 spectra averaged together.  The inset is the difference between the two and shows the hydrogen induced bonding state.

**Tables**

Table I:
Results of ATLEED analysis for the structure of the NiAl(110) substrate with 0.5 ML adsorbed H. Shown are the results for both hydrogen ignored and included in the comparison of theoretical versus experimental integer order *I-V* profiles. $\Delta d_{ij}(X)/d_o$ is the percent change in the layer spacing between plane i and j for element X compared to the bulk value $d_o = 2.041$ Å.

|  | No Hydrogen | Hydrogen Included |
|---|---|---|
| $\Delta d_{12}(Ni)/d_o$ | -4.0±0.7 | -3.6±0.5 |
| $\Delta d_{12}(Al)/d_o$ | +3.4±1.0 | +3.6±1.0 |
| Ripple layer 1 (Å) | 0.15±0.01 | 0.15±0.02 |
| Ripple layer 2 (Å) | 0.02±0.02 | 0.02±0.03 |
| Inner potential | 5.76 eV | 5.67 eV |
| $R_p$ | 0.171 | 0.184 |

Table II:
$R_p$ values for best fit structures with hydrogen in different bonding sites

| Bonding Site | $R_p$ |
|---|---|
| Terminal Ni | 0.24 |
| Terminal Al | 0.26 |
| Bridge -- Ni-Ni | 0.17 |
| Bridge -- Al-Al | 0.25 |
| 3-fold -- Ni-Ni-Al | 0.23 |
| 3-fold -- Al-Al-Ni | 0.23 |

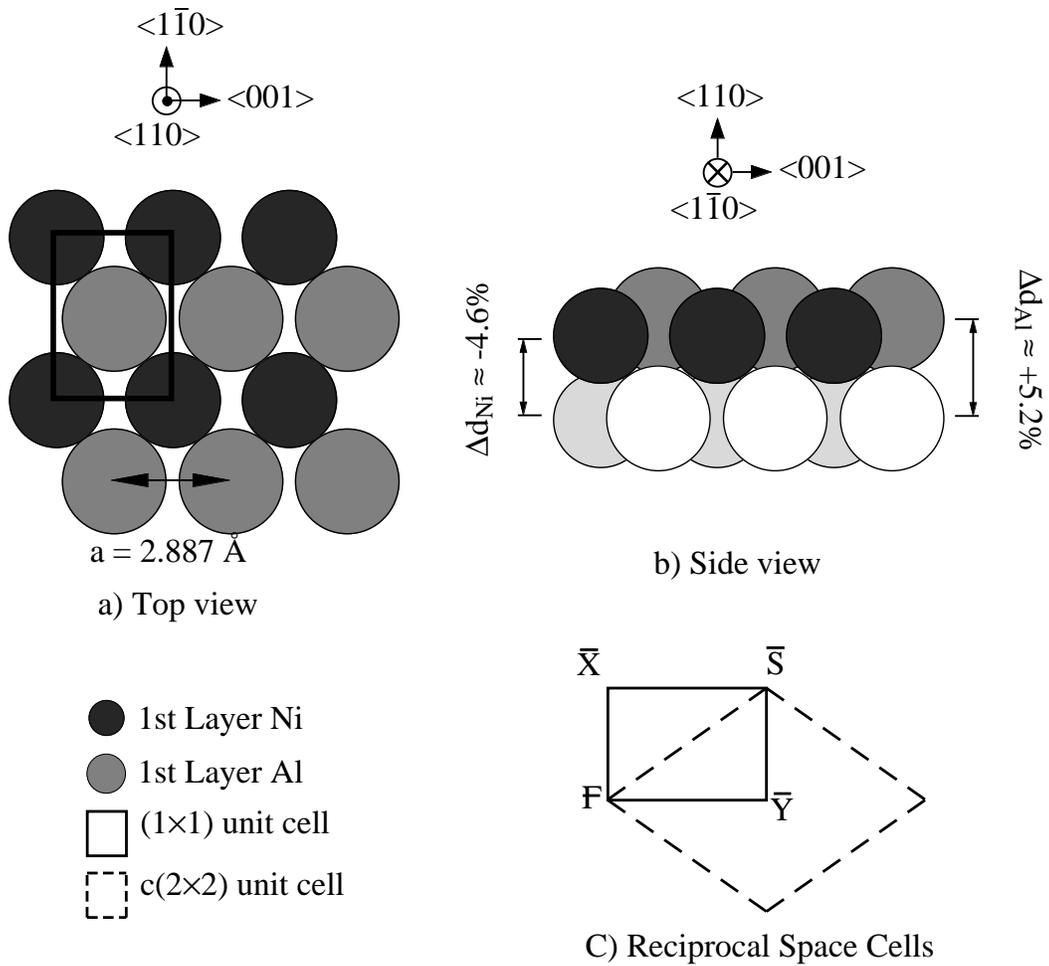

Figure 1: Hanbicki, Rous, and Plummer.

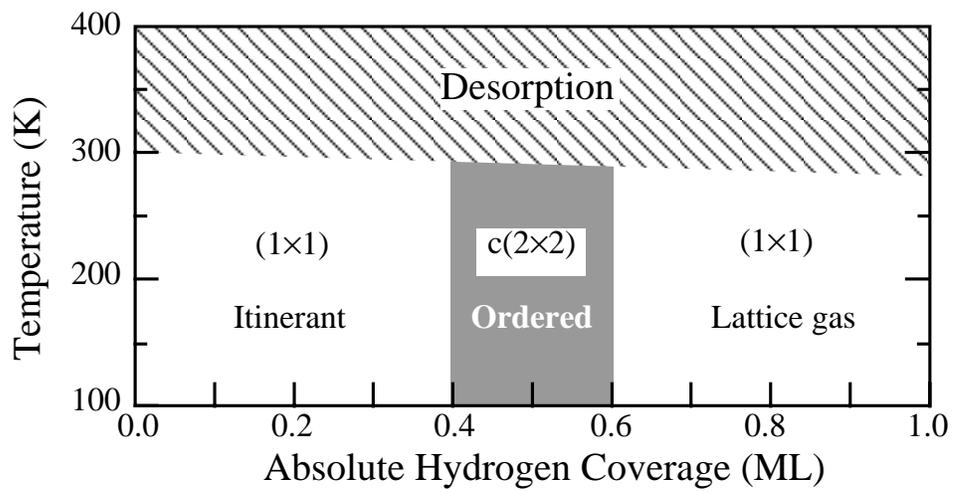

Figure 2: Hanbicki, Rous, and Plummer.

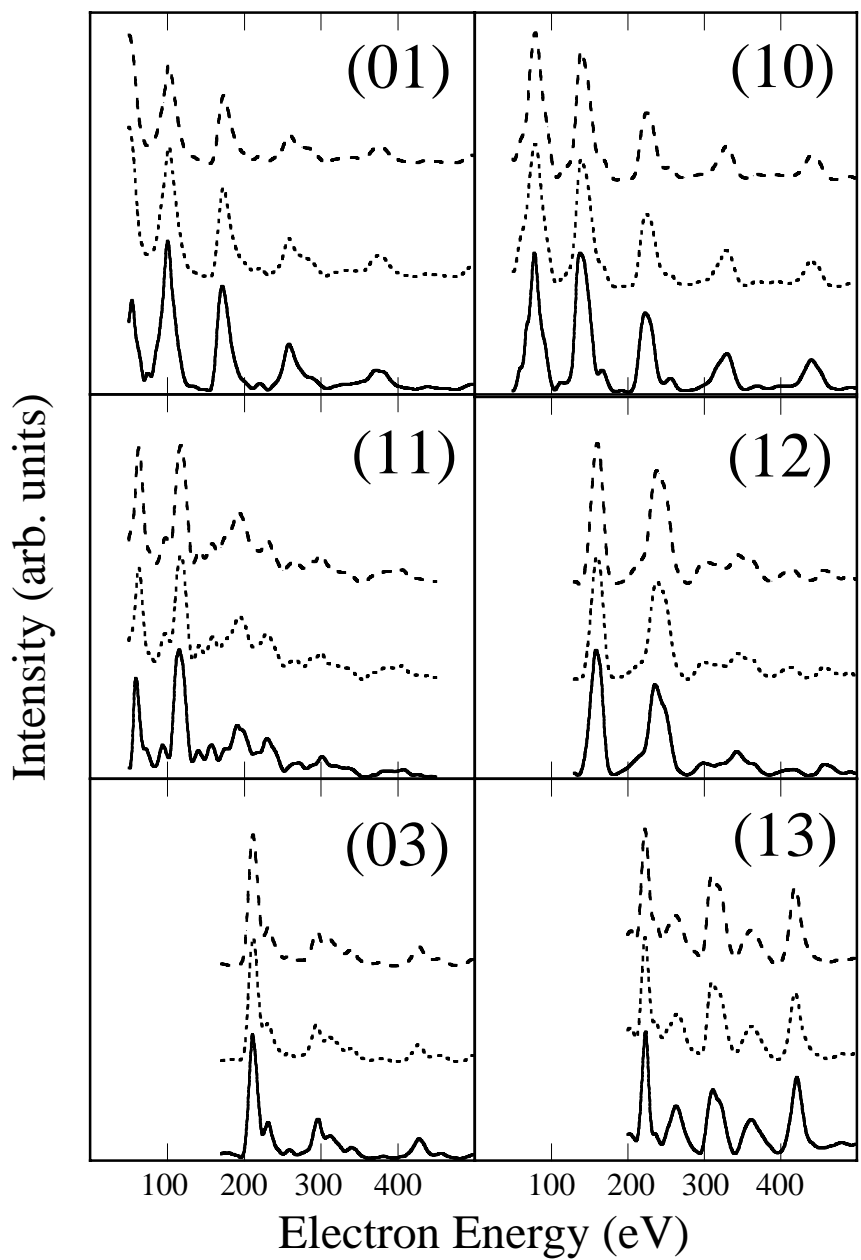

Figure 3: Hanbicki, Rous, and Plummer

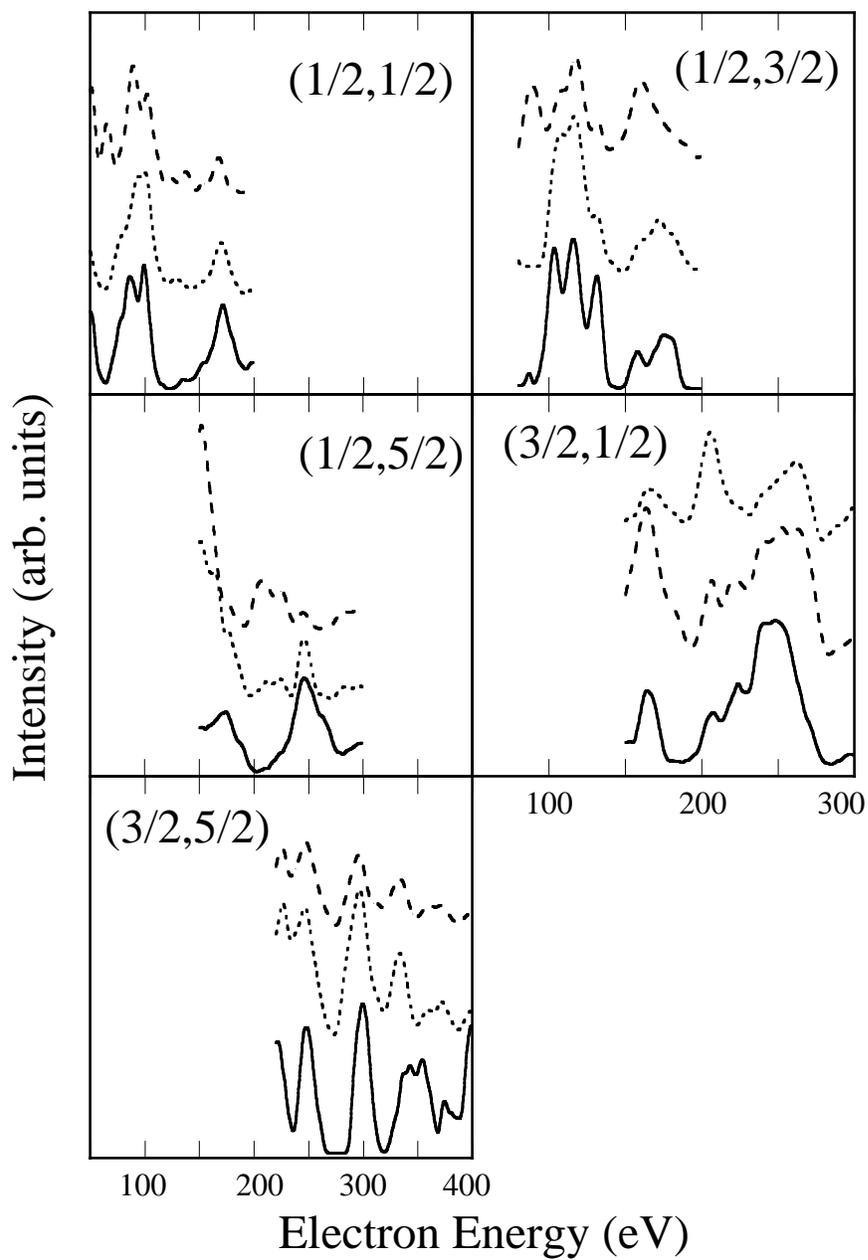

Figure 4: Hanbicki, Rous, and Plummer

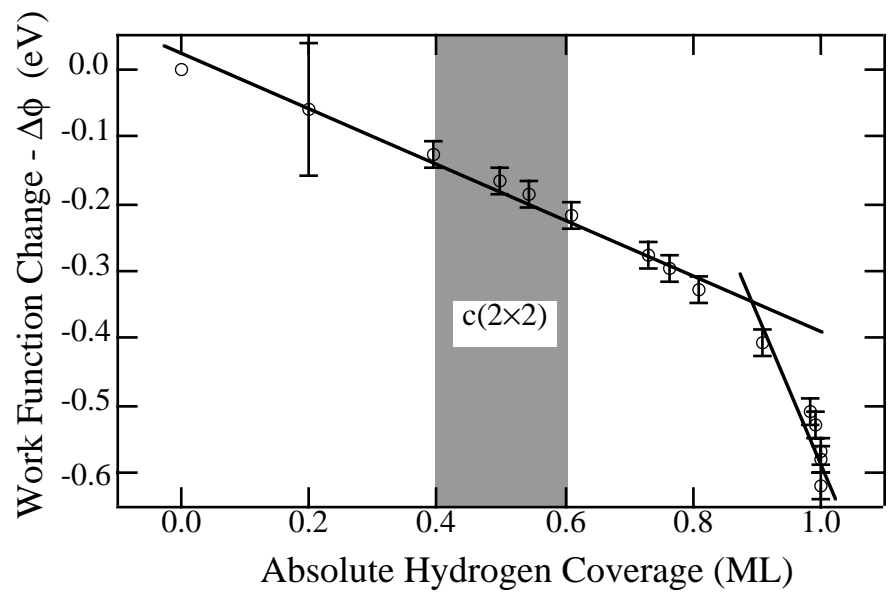

Figure 5: Hanbicki, Rous, and Plummer.

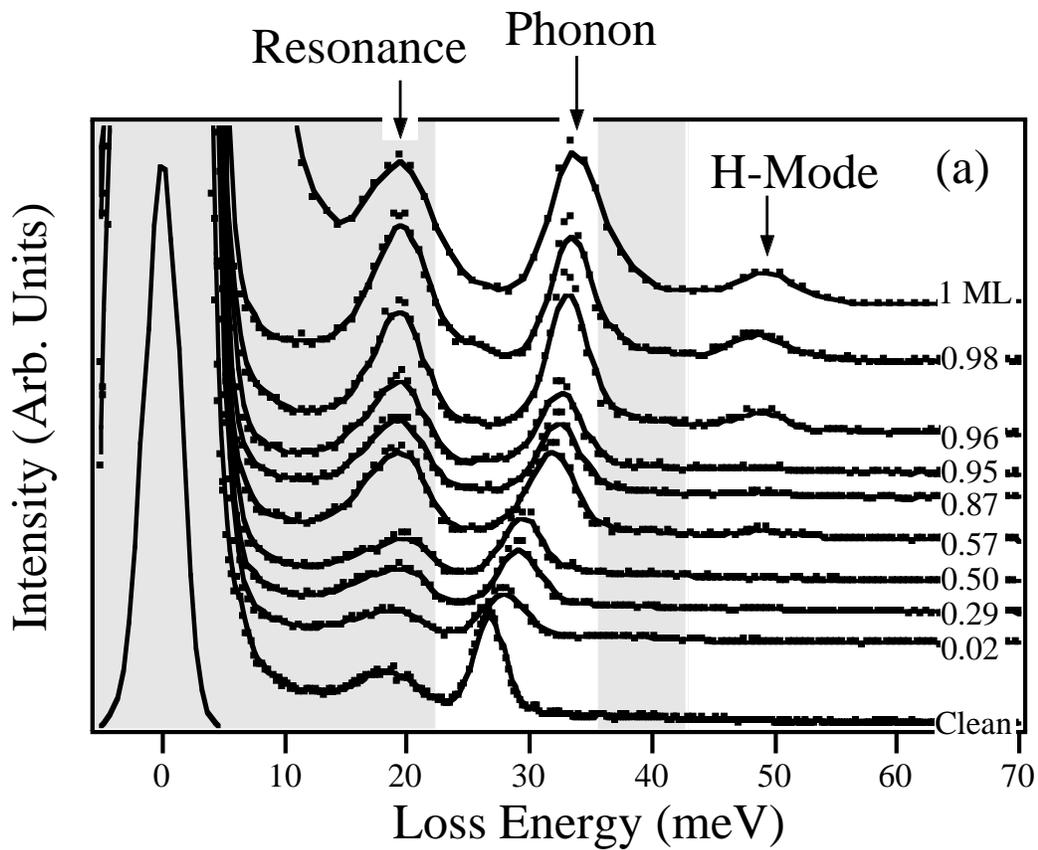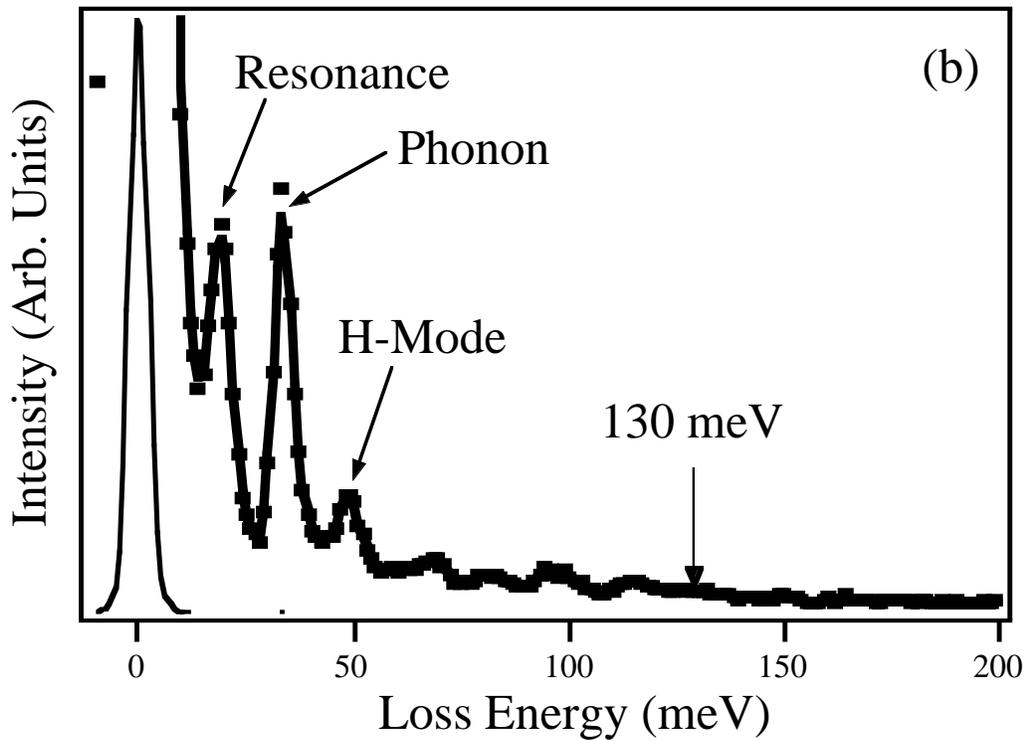

Figure 6: Hanbicki, Rous, and Plummer.

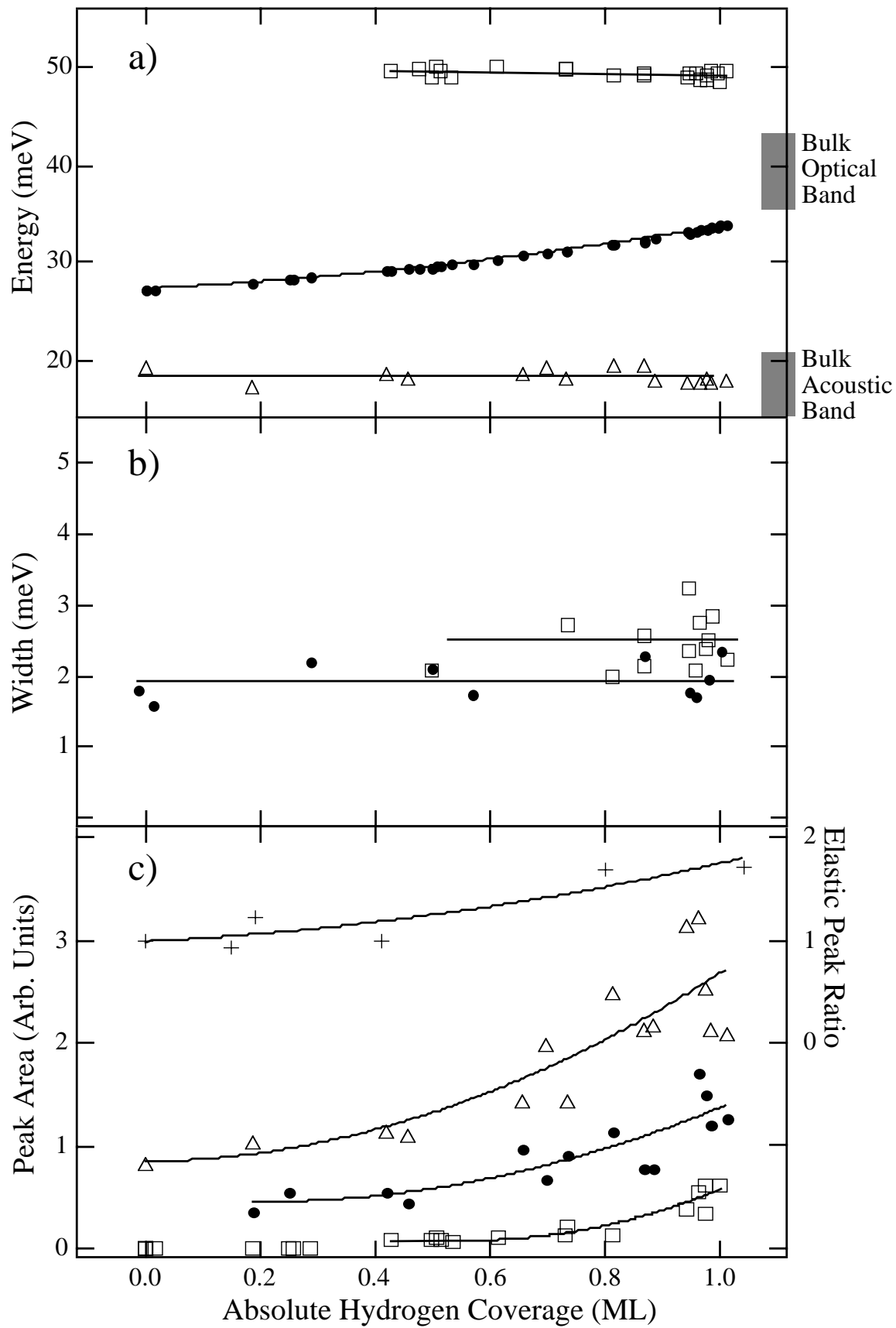

Figure 7: Hanbicki, Rous, and Plumer

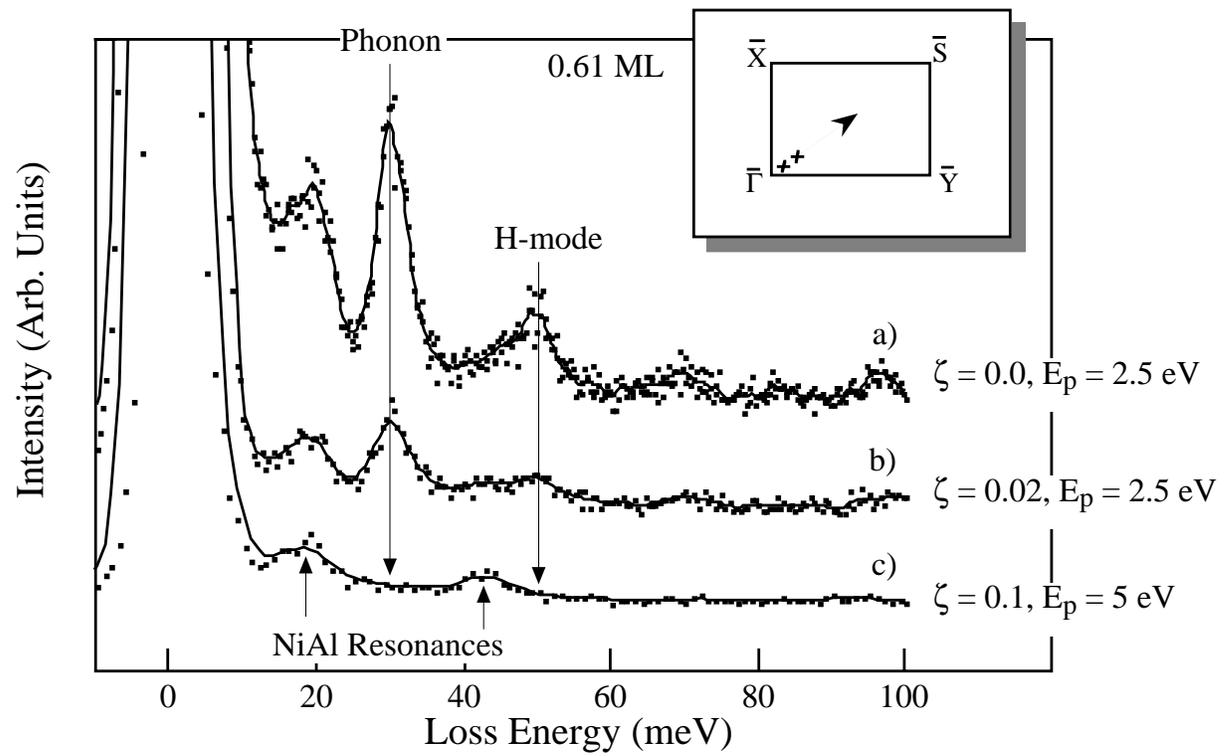

Figure 8: Hanbicki, Rous, and Plummer

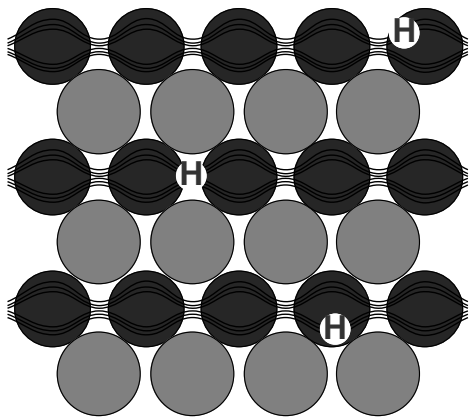
a) Low Coverage -- Itinerant

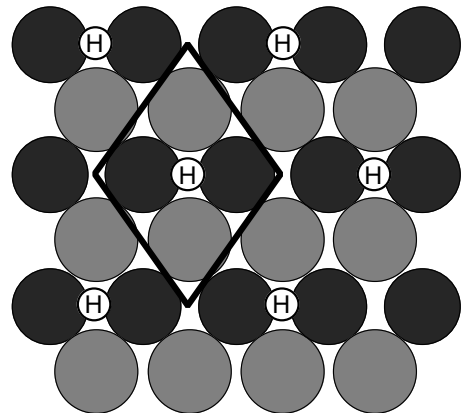
b) 0.5 ML -- Ordered c(2×2)

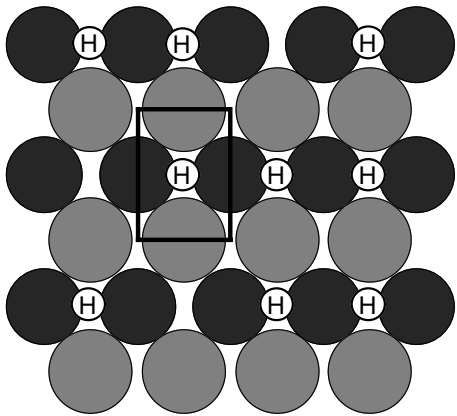
c) 3/4 ML -- Lattice Gas

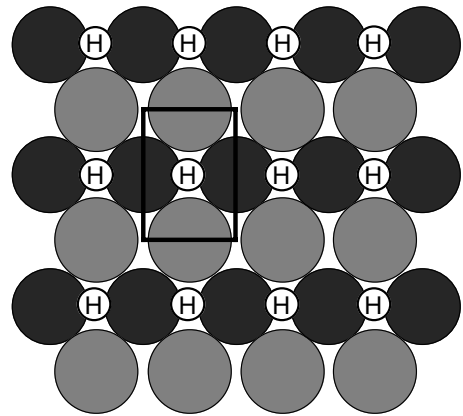
d) 1 ML -- saturation

Figure 9: Hanbicki, Rous, and Plummer.

a)

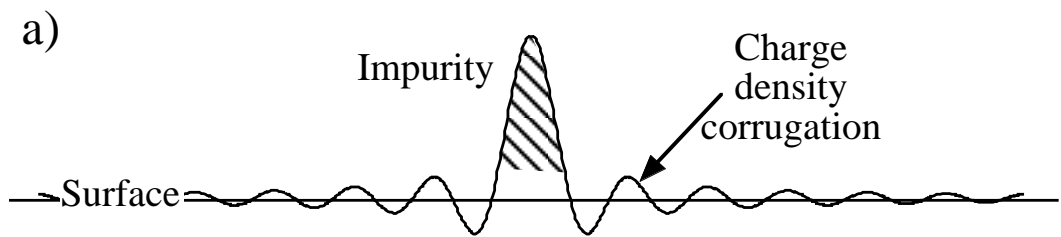

b)

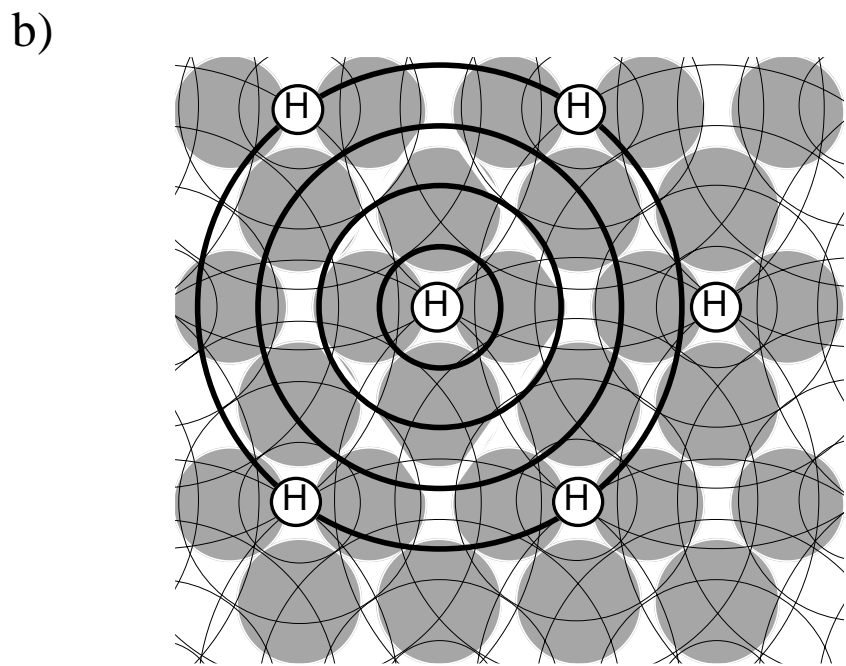

Figure 10: Hanbicki, Rous, and Plummer.

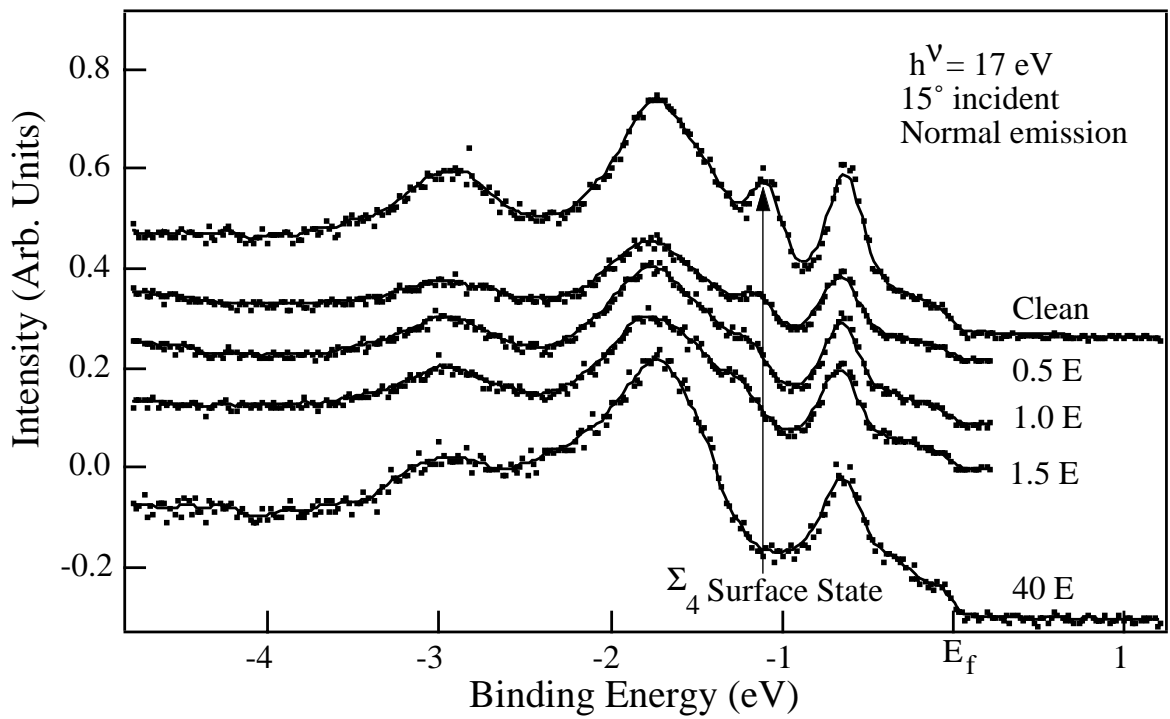

Figure 11: Hanbicki, Rous, and Plummer.

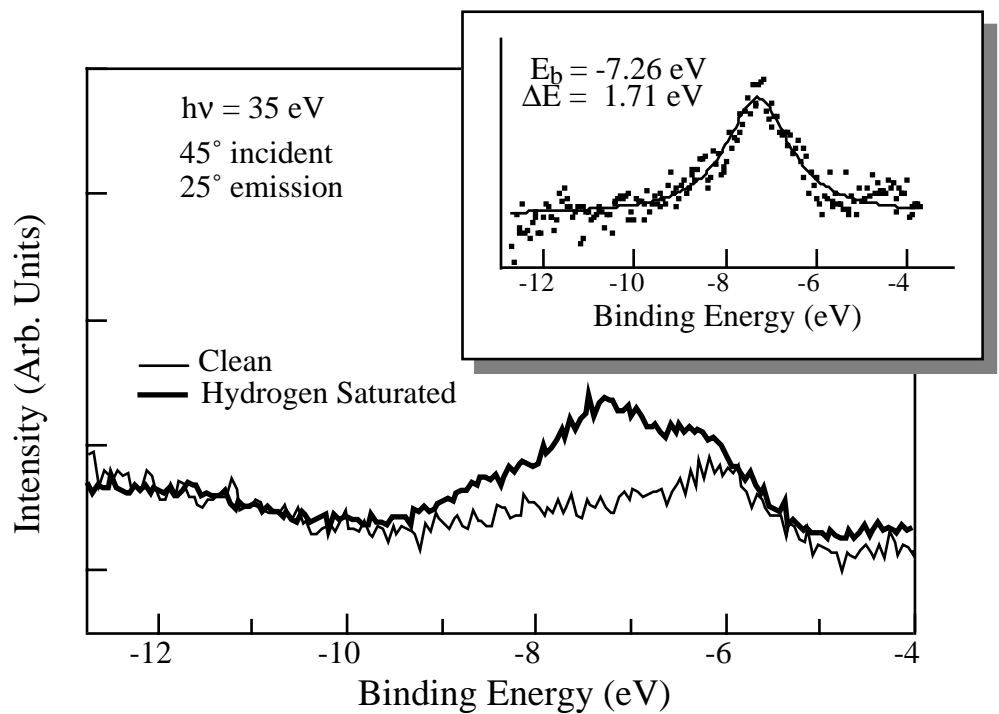

Figure 12: Hanbicki, Rous, and Plummer.